\documentclass[12pt]{iopart}

\usepackage{graphicx}
\usepackage{gensymb}
\usepackage{placeins}
\usepackage{upgreek}
\usepackage[T1]{fontenc}
\usepackage{lineno,hyperref}

\begin{document}

\title[Persisting correlation between electrical and magnetic properties in hexaferrites]{Persisting correlation between electrical transport and magnetic dynamics in M-type hexaferrites}

\author{Željko Rapljenović$^{a}$, Nikolina Novosel$^{a}$, Damir Dominko$^{a}$, Virna Kisiček$^{a}$, David Rivas Góngora$^{a}$, Đuro Drobac$^{a}$, Mladen Prester$^{a}$, Denis Vinnik$^{b}$, Liudmila N.\ Alyabyeva$^{c}$, Boris P.\ Gorshunov$^{c}$, Tomislav Ivek$^{a}$}

\address{$^{a}$Institute of Physics, Bijenička c.\ 46, HR-10000 Zagreb, Croatia}
\address{$^{b}$South Ural State University, 454080 Chelyabinsk, Russia}
\address{$^{c}$Laboratory of Terahertz Spectroscopy, Center for Photonics and 2D Materials, Moscow Institute of Physics and Technology, 141700, Dolgoprudny, Russia}

\ead{zrapljenovic@ifs.hr}
\vspace{10pt}
\begin{indented}
\item[]March 2021
\end{indented}

\begin{abstract}
In this work we present frequency-dependent magnetic susceptibility and dc electric transport properties of three different compositions of hexaferrite Ba$_{1-x}$\-Pb$_{x}$\-Fe$_{12-y}$\-Al$_{y}$\-O$_{19}$. We find a correlation between activation energies of dc electric transport and ac magnetic susceptibility which persists in the whole researched range of aluminium substitution $x=0$ to $3.3$. This result is discussed in the context of charged magnetic domain walls, the pinning of which is determined by charge carriers activated over the transport gap. Our work points toward a general relaxational mechanism in ferrimagnetic semiconductors which directly affects dynamic magnetic properties via electric transport.
\end{abstract}

\section{Introduction}

Hexagonal ferrites (hexaferrites) are widely used and actively studied magnetic materials, which influenced both technological and scientific developments. Although known for more than half a century, they are still a fruitful topic in the research of physical phenomena such as multiferroicity and magnetoelectric effect \cite{HernndezGmez2020,Zhai2020,Kimura2012}. Somewhat recently, it has been proposed that hexaferrites are hosts to the so-called quantum electric dipole liquid, a phase that could be described as the electric analog of the elusive quantum spin-liquid state \cite{Shen2016}, in which electric dipoles persist in a superposition configurations without long-range ordering down to the lowest of temperatures. This kind of exotic state raises the question about the nature of possible magnetoelectric coupling in hexaferrites and the relation with their magnetic order.

\begin{figure}
    \centering
    \includegraphics[width=0.6\textwidth]{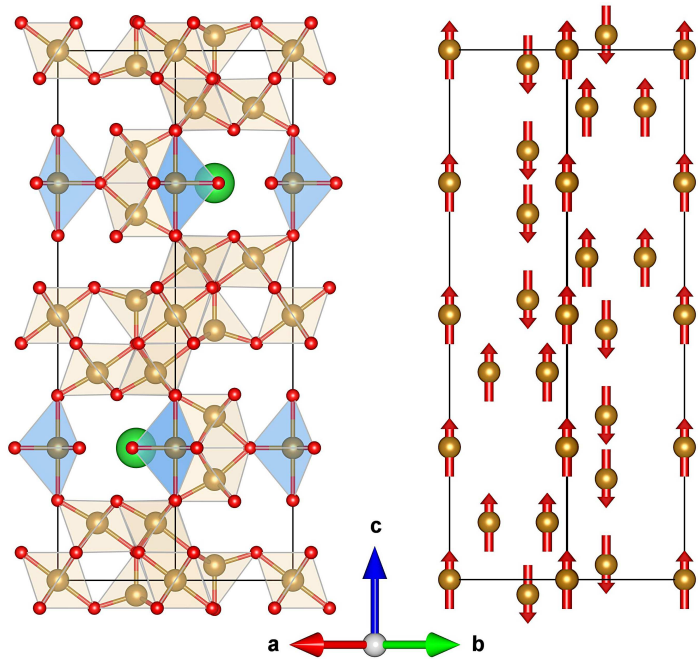}
    \caption{Crystal(left) and magnetic(right) structure of M-type hexaferrite BaFe$_{12}$O$_{19}$ \cite{Momma2011,Jain2013} as viewed along [$\overline{1} \, \overline{1} \,0$] direction. Green spheres represent barium atoms, red represent oxygen, while gold spheres and polyhedra represent iron atoms and their coordinations. Blue hexahedrons mark trigonal by-pyramidal sites. Red arrows represent local iron magnetic moments.}
    \label{fig:1}
\end{figure}

The most famous type of hexaferrites is the so called M-type with the general formula $AB_{12}$O$_{19}$, where $A$ is usually an alkaline earth metal and $B$ is usually iron. A well known member of the M-type hexaferrite class is the BaFe$_{12}$O$_{19}$ (BaM). At room temperature it has a centrosymmetric hexagonal structure, space group $P6_{3}/mmc$. Figure \ref{fig:1} shows the unit cell of BaM which contains 2 formula units \cite{Jain2013}. Fe$^{3+}$ occupy five crystallographically non-equivalent sites in three different oxygen coordinations: octahedral (Oh), tetrahedral (Th) and trigonal bipyramidal (TBP). The magnetic state of this compound is collinear ferrimagnetic up to $T_\mathrm{N} \approx 470\,\degree$C with 16 aligned magnetic moments which are compensated by 8 moments of the opposite spin resulting in $20\mu_{B}$ per formula unit at $T=0$\,K \cite{PULLAR20121191}. Magnetic anisotropy of this system is uniaxial with easy direction being the c-axis, with high values of magnetocrystalline anisotropy constant $K_{1} = 3.25 \times 10^{5}$\,Jm$^{-3}$. The high magnetocrystalline  anisotropy originates in single-ion anisotropy of Fe$^{3+}$ ions in the TBP positions \cite{wohlfarth1980ferromagnetic,Anizotropija-bipiramida}. Interestingly, apart from localized spins BaM also features activated carriers and is a small-gap semiconductor, with its conduction mechanism due to hopping between Fe$^{2+}$ ions \cite{Otpor}. 

The exact positions of iron atoms within TBP coordination and their temperature evolution is still under discussion due to their effect on the overall symmetry. If the iron atom resides in the equatorial plane formed by the three surrounding oxygen sites, as in Figure \ref{fig:1}, the structure is centrosymmetric. If iron is displaced towards one of the the apical oxygen atoms, the structure is polar and with it a permanent electric dipole moment is allowed. Neutron diffraction performed at room temperature showed that iron atoms are shifted towards one of the apical oxygens by $0.26$\,\AA{}, while upon cooling to 4.2\,K the iron atoms seemingly freeze in the equatorial plane \cite{Collomb1986}. The low-temperature result was additionally confirmed by the phonon spectrum calculation \cite{PhysRevB.44.619}. The dielectric measurements on BaM showed that the dielectric function follows the Barret formula from room temperature down to 2\,K which is indicative of quantum paraelectric behavior \cite{Shen2016,Shen2014}. This is likely caused by the polar distortion of TBP sites (somewhat contradicting the neutron diffraction results) which interact predominantly antiferroelectrically. Since these sites on their own form a triangular structure, all the prerequisites are met for the abovementioned quantum electric dipole liquid.

A recent investigation of (Ba, Pb) chemical substitution in BaM found slow relaxational excitations in dielectric as well as magnetic response with very similar mean relaxation times and shared values of activation energy \cite{Alyabyeva2019}. This so-called birelaxor-like behavior is most likely due to magnetic domain wall dynamics and points toward possible magnetoelectric effect in M-type hexaferrites where TBP iron sites might play a role in coupling of spin and electric dipoles. It was unclear whether the proposed behaviour was specific for one particular composition or it could be generalized to a wider range of substitutions. With that in mind, we searched for evidence of common activation energies in magnetic and electric transport properties of hexaferrites with different compositions. For that purpose in this work we substitute iron with isovalent aluminium in single-crystal samples of the general formula (Ba,Pb)(Fe,Al)$_{12}$O$_{19}$ and examine their electric transport, ac magnetic susceptibility, and the mutual influence of electrical and magnetic properties at low temperatures.

\section{Experimental methods}
In order to exclude extrinsic effects common to polycrystalline samples, we use high quality M-type barium hexaferrite single crystals of Ba$_{0.3}$\-Pb$_{0.7}$\-Fe$_{12}$\-O$_{19}$, Ba$_{0.2}$\-Pb$_{0.8}$\-Fe$_{10.8}$\-Al$_{1.2}$\-O$_{19}$, and Ba$_{0.2}$\-Pb$_{0.8}$\-Fe$_{8.7}$\-Al$_{3.3}$\-O$_{19}$. The samples with characteristic hexagonal faceting of up to 5\,mm  were grown by flux technique using a resistive furnace with a precise temperature control \cite{Alyabyeva2019,Vinnik2014-samples}. For the exploration of magnetic and electronic properties of hexaferrites we used two experimental techniques: dc resistivity and ac magnetic susceptibility measurements.

\paragraph{DC resistivity}Single crystals were measured using standard co-linear 4-probe configuration and 2-probe technique in high resistance regime. Conductive silver paste DuPont 4929N was used to paint contacts in the $\mathbf{E}||c$ orientation. DC electric transport measurements were performed in temperature range from 4.2\,K to 300\,K.

\paragraph{AC magnetic susceptibility} 
Dynamic magnetic susceptibility properties were measured on CryoBIND and homemade high-precision ac susceptibility measurement systems at frequencies 10\,Hz -- 10\,kHz and at temperatures from 4.2\,K to 400\,K. In order to avoid any nonlinear effects inherent to soft ferro- and ferrimagnetic systems, low excitation amplitudes of 25--100\,mOe were employed and Earth's magnetic field was compensated. Homemade ac susceptibility measurement system is additionally equipped with electrical contacts which provide dc electric field in sample environment.

\section{Results}
\subsection{Magnetic relaxation}
\begin{figure}
    \centering
    \includegraphics[width=1.0\textwidth]{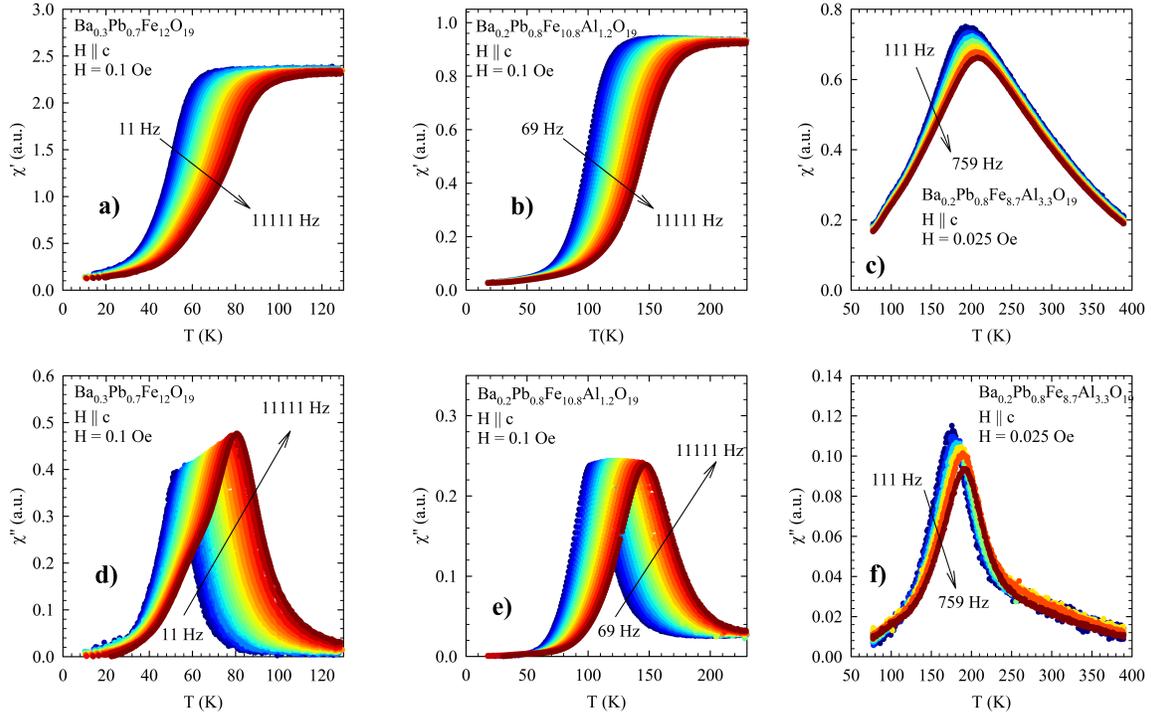}
    \caption{AC susceptibility for different compositions of (Ba,Pb)(Fe,Al)O$_{19}$ M-type hexaferrites along the easy axis $\mathbf{H}||c$ as a function of temperature and frequency. Rows display real [a), b) and c) panels] and imaginary [d), e) and f) panels] part of ac susceptibility, while columns correspond to samples Al0 [a) and d) panels], Al1 [b) and e) panels], Al3 [c) and f) panels]. Colors represent different measurement frequencies. Al0 and Al1 samples show broad relaxational behavior most likely due to magnetic domain dynamics. On the other hand, the Al3 results suggest superparamagnetic response.}
    \label{fig:2}
\end{figure}

Temperature dependence of real and imaginary parts of ac magnetic susceptibility $\chi'$, $\chi''$ (in arbitrary units) of the Al0, Al1, Al3 samples are shown in Figure \ref{fig:2}. For all three compositions we find the largest susceptibility with magnetic field parallel to the $c$-axis, meaning the easy axis of these samples is most likely along $c$-axis. The $\mathbf{H}||c$ results for Al0 were published previously by Alyabyeva et al., \cite{Alyabyeva2019}.

\begin{figure}
    \centering
    \includegraphics[width=0.4\textwidth]{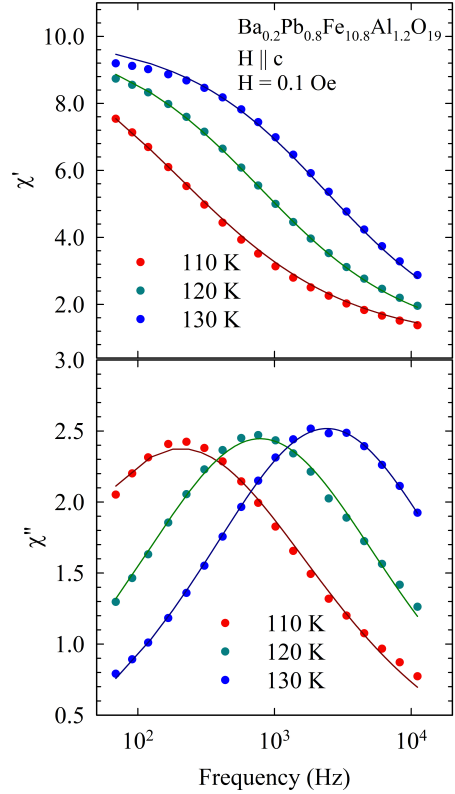}
    \caption{Frequency dependence of the measured ac susceptibility at 110\,K, 120\,K and 130\,K of the Ba$_{0.2}$Pb$_{0.8}$Fe$_{10.8}$Al$_{1.2}$O$_{19}$ (Al1) sample with the corresponding Cole-Cole fit lines (Eq.\ \ref{CCeq}).}
    \label{fig:3}
\end{figure}

The measured susceptibilities have a pronounced frequency dependence which indicates the presence of a relaxation process. Figure \ref{fig:3} shows representative spectra at fixed temperatures for Al1 (Al0 displays a similar behavior, not shown). The broad, step-like real part of susceptibility is accompanied by the bell-like imaginary part. Which suggests a relaxation process is present at low frequencies. The relaxational part is  well described by the Cole-Cole expression
\begin{equation}
\chi'(\omega) - i\chi''(\omega) = \frac{\Delta \chi}{1+ (i\omega \tau_0)^{1-\alpha}} + \chi_\infty,
\label{CCeq}
\end{equation}
where $\Delta \chi$ is the relaxation strength, $\omega = 2\pi f$ is the angular frequency, $\tau_0$ is the mean relaxation time, $\alpha$ represents the broadening of the relaxation time distribution, and $\chi_\infty$ is the high frequency susceptibility. The mean relaxation time corresponds to the maximum of the imaginary susceptibility in frequency domain. We find that fit parameters are not significantly influenced by the high temperature high frequency deviation. Notably, the mean relaxation times $\tau_0$ of Al0 and Al1 samples (shown in Figure 1. of the supplemental) follow Arrhenius type behavior $\tau_{0}\propto e^{E_\mathrm{a} / k_b T}$ with activation energies $E_\mathrm{a} = 1100$\,K and 1750\,K for Al0 and Al1, respectively. Broadening $\alpha$ is constant in the given temperature interval with values $\alpha = 0.61 \pm 0.05 $ for both Al0 and Al1 samples. $\Delta \chi$ is also constant in the temperature interval. The absolute value of this parameter does not directly aid analysis as it strongly depends on size and shape of the sample, for completeness we obtain $\Delta \chi = (1\pm 0.2)\times10^{-2}$ and $(2\pm 0.5)\times10^{-3}$ for Al0 and Al1 samples, respectively. 

The magnetic relaxation in Al0 and Al1 is most likely caused by cooperative spin reversal over the domain walls of the ferrimagnetic long-range order. Regions at and near domain walls are thermally activated over the spin reversal barrier $E_\mathrm{a}$ and subsequently attempt to relax back to a state close to equilibrium. Such a relaxation is typically influenced by the concentration and distribution of pinning impurities, strength of pinning, spin-lattice relaxation, and various other system-dependent properties \cite{Baanda2013,Prester2011}. 
\begin{figure}
    \centering
    \includegraphics[width=0.5\textwidth]{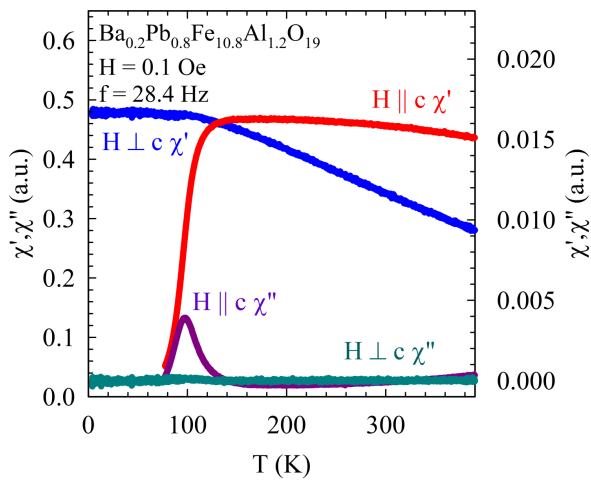}
    \caption{Anisotropy of measured ac susceptibility for the sample Al1. Red and purple curves are real and imaginary susceptibility for $\mathbf{H}\parallel c$ (left y-axis) while blue and green are real and imaginary susceptibilities for $\mathbf{H} \perp c$ (right y-axis).}
    \label{fig:4}
\end{figure}

Sample Al3 displays a relaxation process which appears quite different from Al0 and Al1. A relaxation peak in $\chi''$ {\it vs} $\omega$ is clearly visible only in a very narrow temperature range, which suggests a significantly higher activation energy of $\tau_0$. Furthermore, the peak-like shape of $\chi'$ {\it vs} $T$ resembles that of superparamagnets \cite{Baanda2013}. This means that in Ba$_{0.2}$Pb$_{0.8}$Fe$_{8.7}$Al$_{3.3}$O$_{19}$ the 27.5\% of Fe being substituted by Al is enough to destroy the long-range order of iron spins, i.e., the ferrimagnetic monodomains are confined to small volumes which interact only weakly with other ferrimagnetic monodomains or alternatively this substitution is enough to create almost non-interacting clusters through percolation. For the superparamagnetic processes, mean relaxation time $\tau_0$ is found as maximum of susceptibility in temperature domain for each frequency, due to narrow dispersion of susceptibility with frequency. Interestingly, the superparamagnetic relaxation time of Al3 (shown in Figure 1. of the supplemental) also follows the Arrhenius type behavior with an activation energy $E_\mathrm{a} = 3650$\,K. 

Since M-type hexaferrites are known for their large values of uniaxial anisotropy, for Al1 we additionally performed measurements of ac susceptibility in the $\mathbf{H}\perp c$ direction, ie., perpendicular to the easy axis. The results are shown in Figure \ref{fig:4} and are contrasted with the $\mathbf{H}\parallel c$ response which appears to be about 20 times higher. The response in $\mathbf{H}\perp c$ is functionally different from the $\mathbf{H}\parallel c$ direction in the sense that there is no decrease in real part of ac susceptibility. Rather, a kink (flattening of the blue curve at $100$ K) appears at temperature where the relaxation strength for $\mathbf{H}\parallel c$ is highest. For temperatures below the kink the real susceptibility is temperature independent ($\mathbf{H}\perp c$), meaning that the number of responding entities remains constant. Interestingly, this kink in the $\mathbf{H}\perp c$ direction is frequency-dependent (shown in Figure 3. of the supplemental), and occurs at the temperature where the relaxation is in maximum for the $\mathbf{H}\parallel c$, consistently for every measured frequency. On the other hand, imaginary susceptibility along $\mathbf{H}\perp c$ direction is constant in the whole temperature interval, ie., the phase shift between the excitation and the response does not change. This type of anisotropic response, where susceptibility in the easy direction decreases with temperature and remains constant in the hard direction, is similar to  uniaxial antiferromagnets near $T_\mathrm{N}$ \cite{blundell_magnetism_2014}.



\begin{figure}
    \centering
    \includegraphics[width=0.6\textwidth]{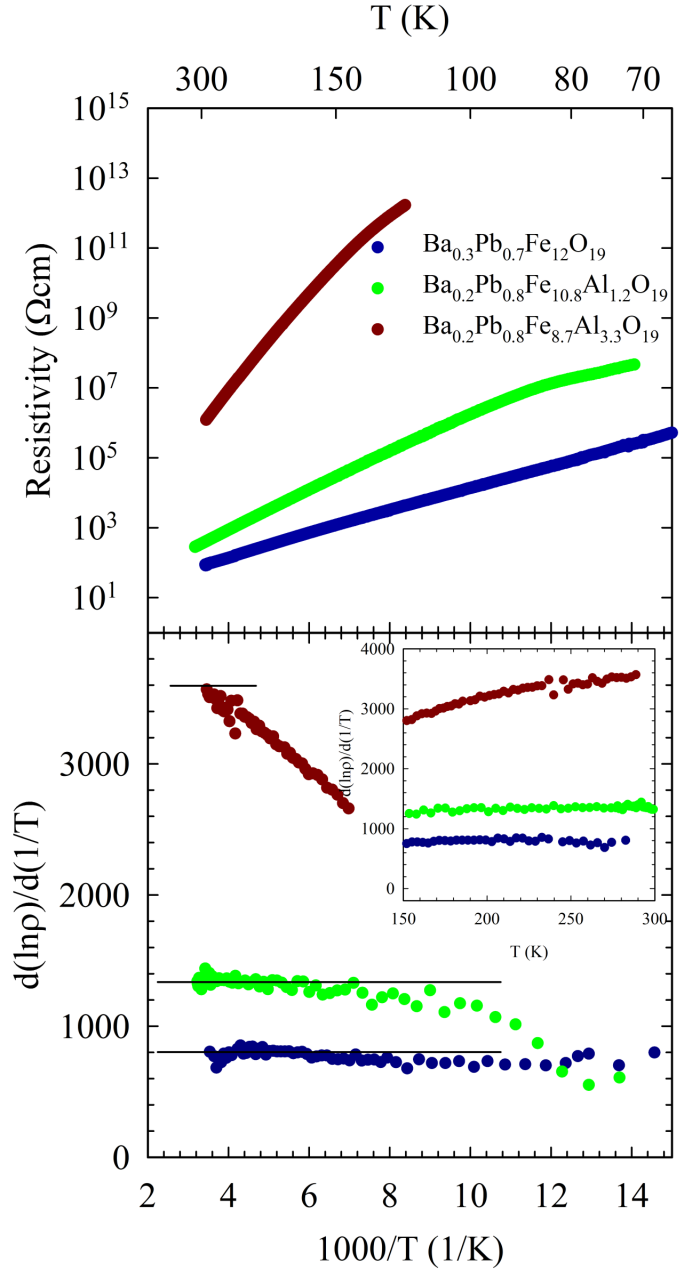}
    \caption{Top plot: dc resistivity of various compositions of (Ba,Pb)(Fe,Al) M-type hexaferrites, $\mathbf{E}\parallel c$. Blue, green and brown line correspond to measured data of Al0, Al1 and Al3 samples, respectively. Bottom plot: Temperature dependence ($1000/$T) of activation energies obtained from resistivity (upper plot) where black lines indicate extracted activation energies for every sample. Inset in the bottom plot is the the temperature dependence of the logarithmic derivative in the linear temperature scale. 
    }
    \label{fig:5}
\end{figure}
\subsection{Activated behavior of dc resistivity}
In the upper panel of Figure \ref{fig:5}, Arrhenius plots of dc resistivity are shown for the Al0, Al1, Al3 samples in $\mathbf{E}\parallel c$ direction. All three systems display semiconducting behavior of the form
\begin{equation}
\label{E:1}
\rho (T) \propto e^{\frac{E_\mathrm{a}}{T}}.
\end{equation}
where $\rho (T)$ is resistivity at temperature $T$ and $E_\mathrm{a}$ is the transport activation energy.  We can estimate the temperature-dependent transport activation energy by noting that $E_\mathrm{a} = \mathrm{d}\ln{\rho} / \mathrm{d}(1/T)$. Shown in the bottom plot of Figure \ref{fig:5}, we see that activation energies of Al0, Al1, Al3 samples are somewhat temperature-dependent with a well-defined value at high temperatures. The decrease at low temperatures can be ascribed to conduction due to localized states within the intrinsic energy gap. Evidently the substitution of iron atoms with isovalent aluminium increases the activation energy.

\section{Discussion}
\begin{figure}[!ht]
    \centering
    \includegraphics[width=0.6\textwidth]{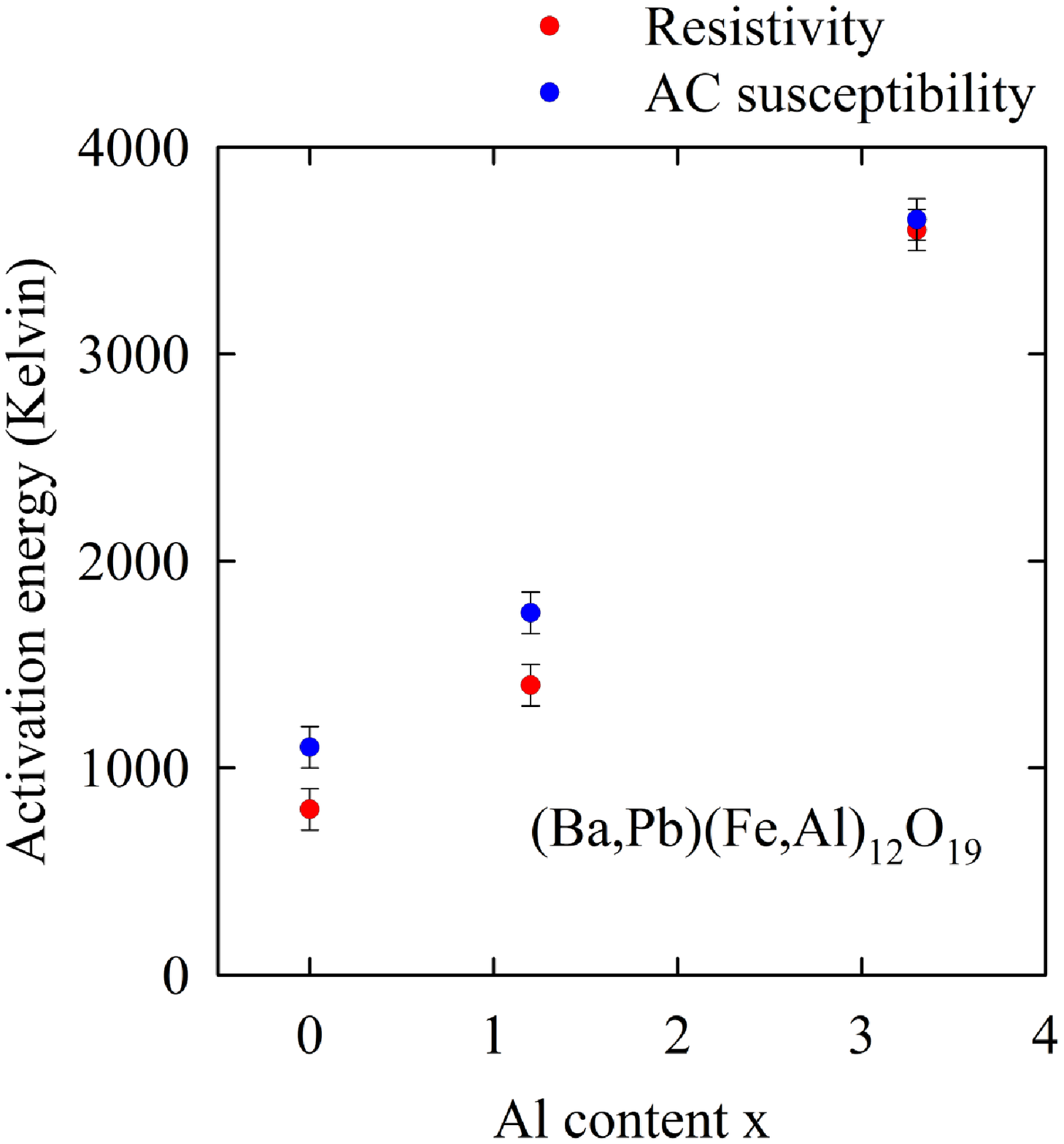}
    \caption{Activation energies obtained from resistivity and ac susceptibility for Al0, Al1, Al3 samples vs Al content.}
    \label{fig:6}
\end{figure}
A comparison of activation energies obtained from dc resistivity and ac magnetic susceptibility measurements is shown in Figure \ref{fig:6}. For all three Al substitutions examined, the two activation energies are very close and show similar increase with the growth of aluminium content. This correlation shows that there could be an underlying process that governs both activation of charge carriers as well as the activated dynamics of magnetic domain walls.

Moreover, the dependence on Al substitution seems to follow a trend. Although the magnetic phenomenology of Al0, Al1, and Al3 compounds is quite different, we may still regard them approximately through the common idea of spin reversal. In superparamagnets, the spin regions being reversed by applied magnetic field are mono-domain particles, while in the long-range-ordered ferrimagnets these can roughly be ascribed to the volume near domain walls. For both situations, the activation energy of spin reversal process is usually taken to be $E_\mathrm{a} = K \cdot V$, ie., proportional to the magnetocrystalline anisotropy energy $K$ and the so-called switching volume $V$ which is roughly the ratio of domain walls in the context of domain wall dynamics or the activation volume from the perspective of superparamagnetic relaxation \cite{Prester2011}. However, this kind of conventional consideration does not take into account the presence of free charge carriers the concentration of which, in the case of Al0, Al1, and Al3, is due to some mechanism thermally activated with an energy close to $E_\mathrm{a}$. Does the presence of free carriers at all relate to magnetocrystalline energy or the switching volume?

The similarity in the activation energies of free carriers and macro-spin reversal barriers invites us to consider magnetoelectric coupling. We performed two experiments (not shown), one where we measured ac susceptibility with applied electric field up to 10\,kV/cm parallel to magnetic field and the another where we measured low frequency dependence of dielectric function (10\,Hz -- 1\,MHz) in magnetic field of 5\,T parallel to electric field. AC susceptibility showed no change with electric field. Also, dielectric function showed no change with application of magnetic field. This does not mean that there is no magnetoelectric effect in the studied hexaferrites (by definition electric polarization induced by magnetic field $P = \alpha H$), rather we found no significant electric/magnetic cross influence in the response up to 10\,kV/cm and 5 \,T. Future work may address the question of magnetoelectric coupling more directly by performing challenging measurements of electric polarization in the presence of free charge carriers eg.\ using techniques discussed by Rivera et al.\ \cite{Rivera2009}.

Our results on (Ba,Pb)(Fe,Al)$_{12}$O$_{19}$ are reminiscent of organic charge transfer salts $\kappa$-(BEDT-TTF)$_2$Cu[N(CN)$_2$]Cl with long-range canted antiferromagnetic order where an analogous form of relaxation was found, albeit in the dielectric subsystem, with an activation energy of mean relaxation times close to the semiconductor transport gap \cite{Tomi__2013,Pinteri1999}. The broad dielectric relaxation was ascribed to the motion of charged domain walls between magnetic domains. The domains are pinned to impurities which are themselves screened by free charge carriers. As temperature was lowered, the concentration of free charge carriers decreased and the weaker screening in turn slowed down the relaxation process of pinned domain walls. Beyond these similarities we show that the hexaferrites possess an anisotropic response similar to antiferromagnets in the region of magnetic relaxation shown in Figure \ref{fig:4}, which extends the similarities between these two systems.  Furthermore, related effects were subsequently found in organic spin liquid $\kappa$-(BEDT-TTF)$_2$Cu$_2$(CN)$_3$ \cite{Pinteric2014} where the relaxor-ferroelectric-like dielectric response can be attributed to charge defects generated within interfaces between frustration-limited magnetic domains, the creation of which is in turn triggered by loss of local inversion symmetry. 

In hexaferrites, recent observations strikingly similar to the organic magnets described above were published by our group for the Al0 single crystals \cite{Alyabyeva2019}. There not only the activation energies of dielectric mean relaxation times and dc resistivity matched, but also corresponded to the activation of mean relaxation times found by ac susceptibility. Relaxor-like dielectric response of Al0 \cite{Alyabyeva2019} implies the existence of electric dipoles, thus the space inversion symmetry must be broken at least locally. We ascribed the birelaxor-like dielectric and magnetic response to the same glassy process of slowing down of magnetic domain dynamics which may carry charge. 

We demonstrate that the correspondence of activation energies found in magnetic relaxation and electric transport indeed applies to a broader range of Al substitution and hence magnetic phenomenology. On this basis we propose a physical picture of magnetoelectric dynamics in which ferrimagnetic domain walls of Al0 and Al1 hexaferrites carry charge or charge dipoles and pin on random charged defects. Here the loss of space inversion symmetry, which is responsible and required for observable dielectric effects, most likely happens at magnetic domain walls where spin inhomogeneities could additionally promote apical displacement of iron ions at the trigonal bipyramidal iron sites. Finally, the strength of screened domain wall pinning is reduced by thermally activated free charge carriers and this predominantly defines the temperature-dependent dispersion of magnetic domain wall dynamics similar to other overdamped, broad relaxor-like dielectrics \cite{Pinteric2014}. We expect that the barriers of macro-spin reversal in superparamagnetic Al3 are similarly lowered by screening via free charge carriers. Indeed this picture is supported by recent magnetic force microscopy and electric force microscopy on single crystal BaM, which found a gradient of electric field across the boundaries of magnetic domains \cite{MFMEFM}.

Finally, we address the origin of charged defects pinning the domain walls. In M-type hexaferrites evidence of Fe$^{2+}$ ions have been found on multiple occasions \cite{Alyabyeva2019,Otpor,8510258}, and its presence could be due to oxygen vacancies formed during sample growth or the lone electrons in Pb 6s orbitals that reduce neighbouring Fe$^{3+}$ to Fe$^{2+}$ as proposed by Alyabyeva et al.\ \cite{Alyabyeva2019}. These Fe$^{2+}$ ions could act as pinning centers for charged domain walls. If so, intentionally introducing Fe$^{2+}$ via aliovalent substitution of iron ions, for example Ti$^{4+}$, could enhance this effect as more pinning centers would become available for the ac magnetic response.   

While this explanation of the magnetic relaxation does seem plausible, additional investigation should be performed on the nature of charged domain walls in hexaferrites. As one alternative explanation, a sublattice of Jahn-Teller centers, if found, could in principle be connected to transport and relaxation of magnetism. Such a sublattice has been identified in Titanium-doped hexaferrite \cite{Gudkov2020} with structural relaxations present in a similar temperature interval and activation energies comparable to our findings. Also, magnetic aftereffect measurements in hexaferrites provide evidence of Fe$^{2+}$ ion diffusion \cite{wohlfarth1980ferromagnetic} which might influence transport via hopping between Fe$^{2+}$ sites \cite{Otpor}. At this point our observations cannot discount the possibility that both Jahn-Teller sublattice and Fe$^{2+}$ ion diffusion may contribute to the appearance of a common activation energy governing magnetic relaxation and semiconducting transport.

In order to clarify the open questions, for future work we propose a set of imaging experiments on Aluminium substituted hexaferrites that are sensitive to both magnetic and electric loacl fields, eg.\ from the class of scanning probe microscopy techniques, which could reveal whether magnetic domain walls carry charge and pin to charged defects. Furthermore, it appears that computational efforts based on density functional theory might shed some light on the slow, long-wavelength dielectric response of hexaferrites by examining the related subtle changes of charge density around spin inhomogeneities at ferrimagnetic domain walls.

\FloatBarrier
\section{Conclusion}
We report on ac magnetic relaxation dynamics and dc electric transport of ferrimagnetic, semiconducting hexaferrites (Ba,Pb)(Fe,Al)$_{12}$O$_{19}$. For three distinct Al substitutions spanning long-range ordered and superparamagnetic, short-range magnetic states, we find that the activation energies governing magnetic relaxation and electric transport correlate to a large degree. We explain the observed magnetoelectric-like feature of hexaferrites in the picture of charged magnetic domain walls that pin to charged impurities, an effect screened by free charge carriers. The impact of this finding is relevant to the design of novel semiconducting soft-ferro- and ferrimagnets where electric transport properties directly affect the dynamics and persistence of magnetic states. Future computational and experimental work is certainly warranted in order to further elucidate the effect of spin inhomogeneities on charge degrees of freedom in these iron-based materials.

\section*{Acknowledgement}
We thank Silvia Tomić for fruitful and enlightening discussions. This work was supported by the Croatian Science Foundation, Grants No.\ IP-2018-01-2730 and IP-2013-11-1011. Liudmila N.\ Alyabyeva acknowledges support of system of President’s scholarships for young researchers (SP-777.2021.5). Denis Vinnik was supported by Grant of President of Russian Federation for young doctors of science (MD-5612.2021.4).

\section*{Literature}
\bibliography{mybibfile}
\bibliographystyle{unsrt.bst}

\end{document}